\documentclass[manuscript,nonacm]{acmart}

\setcopyright{none}
\authorsaddresses{}
\usepackage{booktabs}
\usepackage{graphicx}
\usepackage{subcaption}
\usepackage{tabularx}
\usepackage{caption}

\makeatletter
\let\@mktitle@i\@mktitle@iii
\renewcommand{\@mkauthors@i}{%
  \global\setbox\mktitle@bx=\vbox{%
    \noindent\unvbox\mktitle@bx
    \par
    \begingroup
      \centering
      \normalfont\fontsize{11}{15}\selectfont
      Jiongning Allen Lian\textsuperscript{1}\hspace{1.2em}%
      Yutong Alison Mei\textsuperscript{1}\hspace{1.2em}%
      Ho Ting Cheung\textsuperscript{1}\par
      MK Michael Cheung\textsuperscript{2}\hspace{1.2em}%
      Xiangzhi Eric Wang\textsuperscript{1}\hspace{1.2em}%
      George Baciu\textsuperscript{1}\par
      Henry Been-Lirn Duh\textsuperscript{3}\hspace{1.2em}%
      Qing Li\textsuperscript{1}\hspace{1.2em}%
      Ou Yang\textsuperscript{4}\hspace{1.2em}%
      Zackary P. T. Sin\textsuperscript{1}\par
      \medskip
      \fontsize{9}{12}\selectfont
      \textsuperscript{1}Department of Computing,
      The Hong Kong Polytechnic University, Hong Kong, China\par
      \textsuperscript{2}School of Optometry,
      The Hong Kong Polytechnic University, Hong Kong, China\par
      \textsuperscript{3}School of Design,
      The Hong Kong Polytechnic University, Hong Kong, China\par
      \textsuperscript{4}School of Artificial Intelligence,
      Shenzhen Polytechnic University, Shenzhen, China\par
    \endgroup
    \par\medskip
  }%
}
\makeatother

\begin{document}

\title{Look Over There! Dynamic Visual Guidance for AI-Narrated Slides}

\author{Jiongning Allen Lian}
\affiliation{%
  \department{Department of Computing}
  \institution{The Hong Kong Polytechnic University}
  \city{Hong Kong}
  \country{China}
}

\author{Yutong Alison Mei}
\affiliation{%
  \department{Department of Computing}
  \institution{The Hong Kong Polytechnic University}
  \city{Hong Kong}
  \country{China}
}

\author{Ho Ting Cheung}
\affiliation{%
  \department{Department of Computing}
  \institution{The Hong Kong Polytechnic University}
  \city{Hong Kong}
  \country{China}
}

\author{MK Michael Cheung}
\affiliation{%
  \department{School of Optometry}
  \institution{The Hong Kong Polytechnic University}
  \city{Hong Kong}
  \country{China}
}

\author{Xiangzhi Eric Wang}
\affiliation{%
  \department{Department of Computing}
  \institution{The Hong Kong Polytechnic University}
  \city{Hong Kong}
  \country{China}
}

\author{George Baciu}
\affiliation{%
  \department{Department of Computing}
  \institution{The Hong Kong Polytechnic University}
  \city{Hong Kong}
  \country{China}
}

\author{Henry Been-Lirn Duh}
\affiliation{%
  \department{School of Design}
  \institution{The Hong Kong Polytechnic University}
  \city{Hong Kong}
  \country{China}
}

\author{Qing Li}
\affiliation{%
  \department{Department of Computing}
  \institution{The Hong Kong Polytechnic University}
  \city{Hong Kong}
  \country{China}
}

\author{Ou Yang}
\affiliation{%
  \department{School of Artificial Intelligence}
  \institution{Shenzhen Polytechnic University}
  \city{Shenzhen}
  \country{China}
}

\author{Zackary P. T. Sin}
\affiliation{%
  \department{Department of Computing}
  \institution{The Hong Kong Polytechnic University}
  \city{Hong Kong}
  \country{China}
}

\renewcommand{\shortauthors}{Lian et al.}

\begin{teaserfigure}
  \centering
  \includegraphics[width=\textwidth]{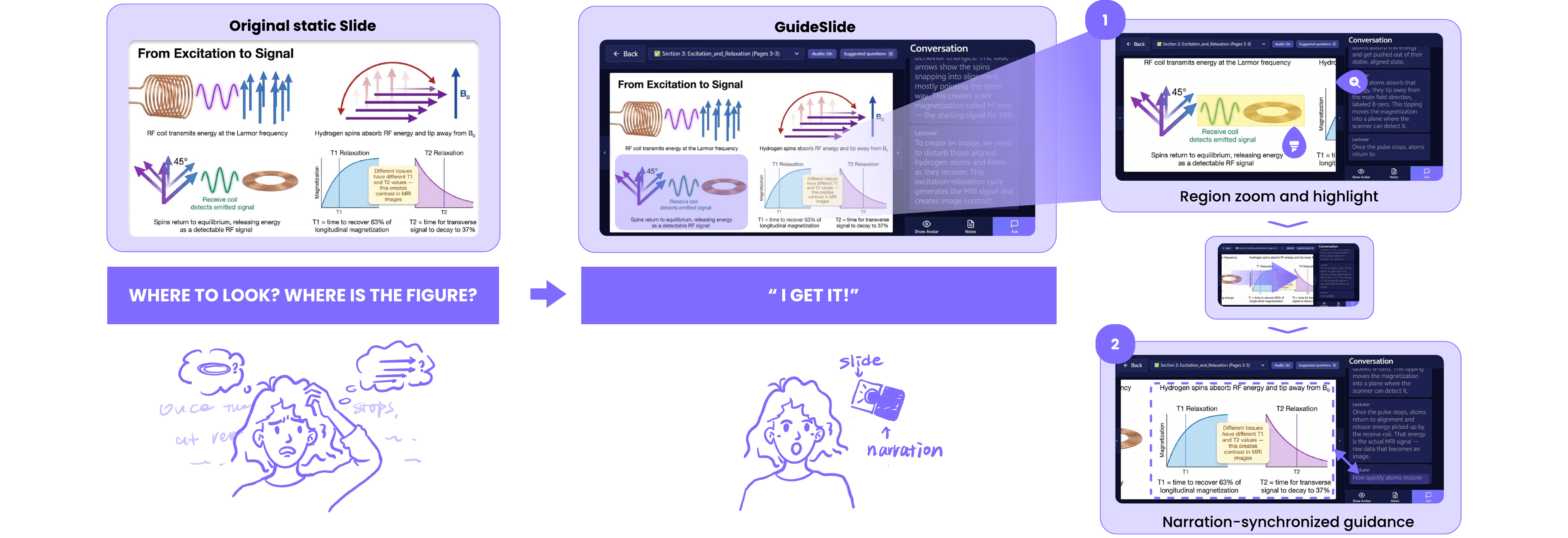}
  \caption{Studying slides without guidance (left) versus with GuideSlide (right). Without guidance, viewers must independently search a dense slide layout for relevant content while processing narration. GuideSlide automatically zooms to the narrated region and applies localized highlights (1), then shifts guidance as narration progresses (2), reducing visual search and split attention throughout the viewing session.}
  \Description{Left: a static educational slide titled ``From Excitation to Signal'' containing four dense diagram panels covering RF coil transmission, hydrogen spin excitation, signal reception, and T1/T2 relaxation curves. A purple banner below reads ``Where to look?'', indicating the absence of navigational guidance. Right: two numbered states of the GuideSlide playback interface. In state~1, the viewport has zoomed into the lower-left panel showing the 45-degree spin diagram and receive coil illustration, with a yellow highlight overlay marking the signal and the coil synchronized to the spoken narration. In state~2, the viewport has shifted to the right panel showing T1 and T2 relaxation curves. A horizontal arrow spanning the two states is labeled ``Visual guidance automatically synchronizes with narration.''}
  \label{fig:teaser}
\end{teaserfigure}

\begin{abstract}
Slides present information spatially, but narration unfolds linearly, leaving the viewer struggling to locate what the narration refers to. This split attention between audio and visual channels forces viewers to search the slide manually, imposing extraneous cognitive load and disrupting comprehension. The problem is especially acute on mobile devices, where dense layouts compressed onto small screens make visual search particularly effortful. In teacher-led settings, however, instructors naturally direct attention through pointing, pacing, and emphasis. Inspired by this, we present \textsc{GuideSlide}, a system that automatically generates \emph{Dynamic Visual Guidance} for slide presentations, providing viewers with narration-synchronized visual cues that direct attention in real time. Given a PDF slide deck, \textsc{GuideSlide} automatically produces region-level zoom and time-aligned highlights without manual annotation. In a within-subjects study with 24 participants, \textsc{GuideSlide} substantially reduced perceived cognitive load and yielded higher immediate comprehension scores compared to static slides with identical narration. Participants strongly preferred the Dynamic Visual Guidance condition, reporting that it made narration easier to follow and reduced uncertainty about where to look. These findings suggest that narration-synchronized visual guidance is a valuable design strategy for slide-based presentations, with potential to extend to other instructional and presentation scenarios.

\end{abstract}



\maketitle

\section{Introduction}
Slides have become one of the most widely used media for structured communication, spanning corporate presentations, academic lectures, training materials, and online courses~\cite{seidel_interaction_2015, zhang-li_awaking_2025, chen_effects_2015}. Yet when viewed without a presenter, slides alone can be difficult to follow~\cite{zhang-li_awaking_2025}. AI-generated narration has emerged as a scalable means of addressing this, automatically providing spoken explanation for slide content~\cite{holmberg2025generatingnarratedlecturevideos, mizukami2025study}. However, narration alone does not resolve a fundamental perceptual challenge: narration unfolds linearly, but the content it refers to is spread across the slide~\cite{ayres_split-attention_2014, mayer_cambridge_2014, chen_effects_2015, clark_dual_1991, mayer_principles_2014}.

Without guidance, viewers are left struggling to locate what the narration refers to~\cite{emhardt_what_2022, jarodzka_learning_2013}. This disconnect between audio and visual channels forces viewers to search the slide manually---a textbook instance of the split-attention effect~\cite{ayres_split-attention_2014, chandler_split-attention_1992}. Visual search itself demands heavy use of working memory. While scanning the slide, viewers must at the same time hold the narration in mind, remember what they are looking for, and evaluate each visual element they encounter---all drawing on the same limited cognitive resource~\cite{baddeley_is_2001, sweller_cognitive_1988}. This leaves fewer resources for actually understanding the content~\cite{mayer_cambridge_2014, sweller_cognitive_1988}. Existing systems that generate narrated slides have attempted to address attention through element-level cues such as highlights or pointer overlays~\cite{holmberg2025generatingnarratedlecturevideos, mizukami2025study}, yet these approaches presuppose that viewers can already locate the relevant region---they signal what to attend to, but not where to look first. Further, these approaches have not been validated with real users.

We can, however, look to human teachers for inspiration. In teacher-led settings, effective attentional guidance often unfolds hierarchically across multiple levels of spatial granularity. Speakers typically use broad cues---such as general gestures, gaze direction, or verbal framing~\cite{li2019gestures, frischen_gaze_2007, van_wermeskerken_seeing_2017}---to establish a spatial anchor that tells viewers where to look. Within this anchored region, they then employ precise pointing to guide viewers through specific elements~\cite{emhardt_what_2022, sauter_distance_2022}. This region-level framing is grounded in the signaling principle~\cite{wagner_exploring_2023, mayer_nine_2003, alpizar_meta-analysis_2020}. By narrowing the visual search space, it reduces the cognitive cost of integrating spoken narration with the corresponding visual content. Yet existing systems address only the second stage, marking which element to attend to, while leaving the first stage unsolved: which region to look at.

Zoom is uniquely suited to address this first stage. Unlike highlights, which mark a specific element but leave the viewer to locate it within the full slide, zoom restructures the visual field itself: by magnifying the target region, it reduces the space the viewer needs to search and brings the relevant content into immediate focus~\cite{plumlee_zooming_2006, amadieu_attention-guiding_2011}. The zooming animation further aids this process by preserving spatial relationships and reducing disorientation~\cite{shanmugasundaram_effect_2008}. As the new region comes into view, this dynamic motion naturally captures the viewer's bottom-up visual attention~\cite{das_shifting_2024, yang_unveiling_2026}, ensuring a seamless transition before micro-level highlights are applied.

Inspired by this two-stage structure, we present \textsc{GuideSlide}, a system that generates \emph{Dynamic Visual Guidance} for slide presentations without manual annotation. Given a PDF slide deck, \textsc{GuideSlide} first zooms to the narrated region to establish a spatial frame, then overlays time-aligned highlights to guide element-level attention---replicating the hierarchical attentional guidance that teachers provide naturally. We evaluate \textsc{GuideSlide} through a within-subjects user study with 24 participants, providing the first empirical evidence that hierarchical guidance combining region-level zoom and element-level highlighting can meaningfully reduce split-attention and improve comprehension.

This paper makes three contributions: (1)~we identify a two-level visual search burden as the core challenge in narrated slide viewing and provide a theoretical account of why macro-level spatial guidance is essential for reducing split-attention; (2)~we design and implement \textsc{GuideSlide}, a fully automatic pipeline that generates hierarchical Dynamic Visual Guidance from arbitrary slide decks without manual annotation, in which zoom serves as an essential layer for macro-level spatial guidance;  (3)~we present a within-subjects user study ($N = 24$) providing the first empirical evidence that hierarchical guidance combining region-level zoom and element-level highlighting can meaningfully reduce split-attention and improve comprehension. We believe \textsc{GuideSlide} is broadly applicable across domains, and particularly valuable in business communication and teaching contexts.

\section{Background and Related Work}

\subsection{The Split-Attention Effect in Narrated Slide Viewing}
The perceptual challenge of narrated slide viewing is well-grounded in cognitive load theory. The split-attention effect~\cite{ayres_split-attention_2014, chandler_split-attention_1992, ginns_integrating_2006, cierniak_explaining_2009} predicts that when related sources of information are spatially or temporally separated~\cite{mayer_principles_2014, schroeder_spatial_2018}, viewers must mentally integrate them at the cost of working memory capacity~\cite{baddeley_is_2001, moreno_cognitive_1999}, generating extraneous cognitive load that leaves insufficient resources for comprehension~\cite{sweller_cognitive_1988, mayer_cambridge_2014}. Empirical evidence further confirms this cost: Chen and Wu~\cite{chen_effects_2015} found that spatially separating instructor video from slide content produced significantly higher cognitive load than integrated formats, underscoring how layout fragmentation alone amplifies the perceptual burden on viewers. Foundational work by Jeung et al.~\cite{jeung_role_1997} demonstrated that audio-visual instruction fails to enhance learning under high visual search conditions unless explicit visual indicators are incorporated to coordinate the two streams, directly motivating the need for synchronized visual guidance~\cite{de_koning_learning_2010}. This visual search burden has two layers: viewers must first identify which region of the slide the narration refers to, and then locate the specific element within that region~\cite{wolfe_visual_2020}.

\subsection{Signaling Principle and Micro-Level Highlighting}
The signaling principle holds that visual cues directing attention to relevant elements reduce the integration cost imposed by split attention, thereby improving comprehension~\cite{mayer_cambridge_2014, alpizar_meta-analysis_2020, de_koning_towards_2009, mayer_nine_2003}. This principle operates at two levels: at the macro level, it concerns directing viewers to the right region of the slide; at the micro level, it concerns marking the specific element within that region. Highlighting is the most established implementation of micro-level signaling. Empirical evidence confirms its effectiveness: dynamic highlights capture attention and reduce extraneous cognitive load~\cite{das_shifting_2024}, and recent EEG and eye-tracking work demonstrates that highlighting can directly promote knowledge acquisition by directing attention to crucial areas~\cite{sun_effect_2025}. However, highlighting presupposes that viewers can already locate the relevant region---it reduces search within a region, but does not tell viewers which region to look at in the first place. Macro-level spatial guidance remains unaddressed.

\subsection{Macro-Level Spatial Guidance through Zoom}
Addressing the macro-level challenge requires restructuring the visual field itself, rather than merely marking elements within it. Zooming interfaces achieve this by magnifying a target region, reducing the spatial extent viewers must search and bringing relevant content into immediate focus~\cite{plumlee_zooming_2006}. Critically, this approach has been shown to reduce unnecessary cognitive load by implicitly filtering out peripheral information, allowing viewers to concentrate on the narrated region without being distracted by surrounding content~\cite{amadieu_attention-guiding_2011}. Eye-tracking research further confirms that digital zoom-ins significantly increase gaze coherence and target dwell times, reducing distracting transitions to surrounding content~\cite{glaser_designing_2017}. Beyond static magnification, the zooming animation itself functions as a macro-level signaling cue: animated transitions preserve spatial relationships between views, reducing disorientation as viewers move between regions~\cite{shanmugasundaram_effect_2008}. As a salient dynamic stimulus, the looming motion of a zoom triggers bottom-up visual attention capture significantly faster than static cues~\cite{franconeri_moving_2003, yang_unveiling_2026, das_shifting_2024}, naturally directing the viewer's gaze to the incoming region before any element-level highlighting is applied. Together, zoom and highlight implement the signaling principle as a unified hierarchical guidance system~\cite{de_koning_towards_2009}: zoom resolves the macro-level \emph{where to look}, and highlight resolves the micro-level \emph{what to attend to}.

\subsection{Automatic Narrated Slide Generation}
The closest existing work focuses on automatically generating narrated lecture videos from slide decks. AutoLectures~\cite{holmberg2025generatingnarratedlecturevideos} and Mizukami et al.~\cite{mizukami2025study} have demonstrated the feasibility of using LLMs and OCR to generate time-aligned lecture videos with synchronized visual cues. However, these approaches share two fundamental limitations. First, they provide no macro-level spatial guidance: visual cues are applied at the element level without first directing viewers to the relevant region of the slide~\cite{holmberg2025generatingnarratedlecturevideos, mizukami2025study}. Without zoom to restructure the visual field, element-level cues presuppose that viewers can already locate the relevant region---leaving the first and more fundamental layer of split attention unresolved. Second, both studies rely exclusively on technical evaluation metrics and explicitly acknowledge the absence of evaluation with real viewers~\cite{holmberg2025generatingnarratedlecturevideos, mizukami2025study}, leaving open whether time-aligned visual guidance can effectively reduce split-attention in practice. \textsc{GuideSlide} addresses both limitations.

\section{Design Rationale}
Our system provides two complementary levels of visual guidance, forming a hierarchical attention guidance scheme that moves from macro level (which region to look at) to micro level (which element to look at), synchronized sentence-by-sentence with the spoken narration and replicating the natural coordination of a teacher speaking while directing attention on the board~\cite{van_wermeskerken_seeing_2017, emhardt_what_2022}.

\paragraph{Region Zoom.}
Region zoom automatically centers the viewport on the region currently being discussed, narrowing the viewer's visual search space and eliminating the need for manual navigation~\cite{plumlee_zooming_2006, amadieu_attention-guiding_2011, glaser_designing_2017}. The importance of spatiotemporal guidance has been demonstrated across presentation media: RadarVR~\cite{liu_radarvr_2023} showed that providing viewers with advance spatial and temporal information about where to look significantly improved attention capture in cinematic VR. However, zoom alone is insufficient to orient viewers within the focused region~\cite{seidel_interaction_2015, de_koning_learning_2010}. To preserve spatial context, each zoom target includes at least 30\% padding around the focal content, so that viewers can see the relationship between the focused element and its neighbors. Animated transitions further reinforce the viewer's sense of position within the overall layout~\cite{haverkamp_is_2023, shanmugasundaram_effect_2008, franconeri_moving_2003, yang_unveiling_2026}.

\paragraph{Element Highlight.}
Within the zoomed region, semi-transparent overlays mark the specific visual elements being referenced, implementing the \emph{signaling principle}~\cite{de_koning_towards_2009, mayer_cambridge_2014}. Critically, these highlights are targeted and constrained rather than broadly applied: research has shown that constrained highlighting improves comprehension over unlimited or absent highlighting~\cite{joshi_constrained_2024, de_koning_towards_2009}. Each highlight is bound to a character range in the narration text, appearing when the narration reaches the corresponding phrase and fading when that phrase ends.

\paragraph{Automatic Generation.}
Traditional slide annotation approaches rely on manual teacher markup or post-production video editing, neither of which scales to large volumes of slide content~\cite{holmberg2025generatingnarratedlecturevideos, mizukami2025study}. Our system uses a vision-language model to jointly generate region decomposition, narration text, and spatially bound highlight metadata in a single pass, requiring no human annotation.
\section{Generating Dynamic Visual Guidance with GuideSlide}



\begin{figure*}[t]
  \centering
  \includegraphics[width=\textwidth]{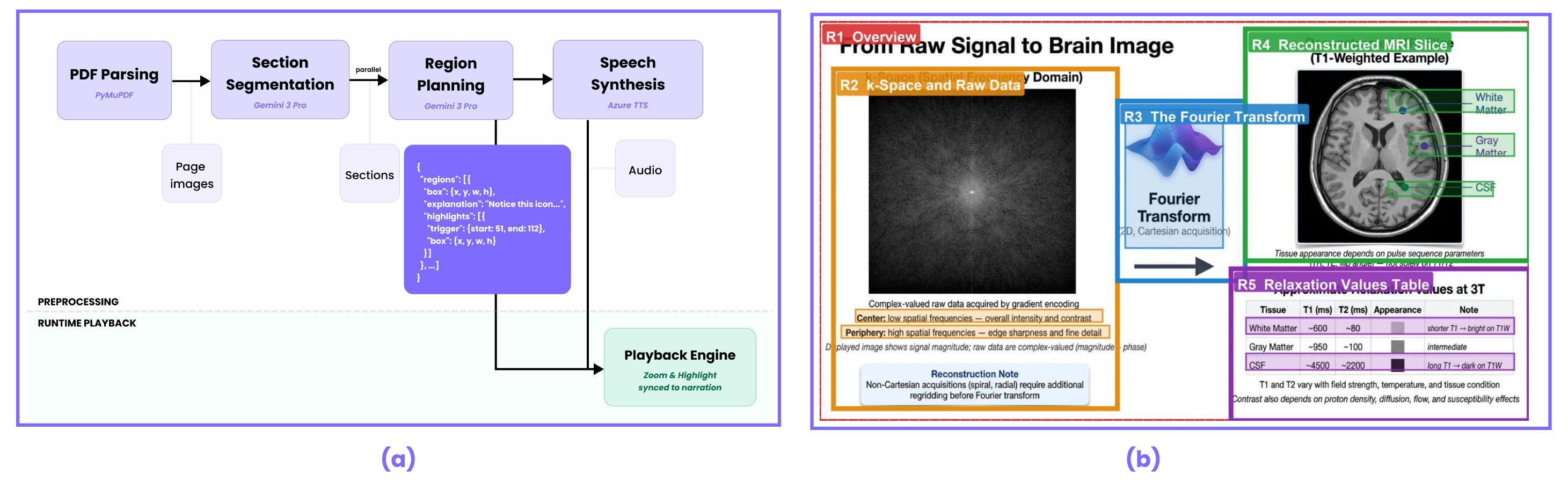}
  \caption{System overview.
    (a)~The generation pipeline and runtime playback: a PDF is parsed into page images, segmented into thematic sections by a vision-language model~(VLM), then each page is independently processed by the VLM to produce structured region data---including zoom bounding boxes, narration text, and character-level highlight triggers---followed by speech synthesis. During playback, these precomputed data are rendered by the playback engine as narration-synchronized zoom and highlight guidance.
    (b)~Region planning output for one slide: color-coded rectangles indicate zoom targets (\textsc{R1}--\textsc{R5}), with numbered labels showing the narration order. Within each region, smaller colored overlays mark element-level highlights whose triggers are each bound to a specific phrase in the narration text.}
  \Description{A two-part figure showing (a) a left-to-right pipeline diagram from PDF input through section segmentation, per-page region planning, and speech synthesis to a playback engine; (b) a sample slide annotated with five color-coded region bounding boxes labeled R1 through R5 in narration order, with smaller colored overlays within regions marking element-level highlights linked to narration text.}
  \label{fig:system-overview}
\end{figure*}

Based on the design goals above, we developed \textsc{GuideSlide}, a system that takes a PDF slide deck as input and automatically generates narration-synchronized region-level zoom and element-level highlights without manual annotation. The system consists of a preprocessing pipeline that generates region plans and speech, and a runtime engine that renders synchronized guidance during playback (Figure~\ref{fig:system-overview}a).

\subsection{Preprocessing Pipeline}


\subsubsection{Section Segmentation.}
The PDF is first converted into page images using PyMuPDF~\cite{pymupdf_docs}. The full deck is then analyzed by a vision-language model (Gemini 3 Pro) to group consecutive slides into thematic sections and assign each section a topic label~\cite{gemini3pro_modelcard}. These section labels are used as context for subsequent processing.

\subsubsection{Region Planning.}
\label{sec:region-planning}
Region planning is the core step of the two-level guidance mechanism: it transforms a static slide page into an ordered set of temporally sequenced narration units mapped to specific slide regions.

\paragraph{Region plan representation.}
For each slide page image, the system generates an ordered region plan $\mathcal{R} = \langle r_1, r_2, \ldots, r_n \rangle$, where each region $r_i = (b_i,\; n_i,\; \mathcal{H}_i)$ contains three components: a \emph{bounding box} $b_i = (x, y, w, h) \in [0,1]^4$ defining the viewport zoom target (region-level guidance); a \emph{narration text}~$n_i$ of two to three sentences rephrasing slide content in a teacher's conversational voice; and a \emph{highlight binding set} $\mathcal{H}_i = \{(s_j,\; e_j,\; \hat{b}_j)\}$, where each binding maps a character range $[s_j, e_j)$ in~$n_i$ to the bounding box~$\hat{b}_j$ of a slide element (element-level guidance). When the narration advances into a trigger range, the corresponding element is highlighted. If no specific element needs to be called out in a region, $\mathcal{H}_i = \varnothing$.

\paragraph{Structural constraints.}

The region plan follows five constraints: (1) an overview-first full-page region; (2) highlight boxes must remain visible within their parent zoom box; (3) each zoom box includes at least 30\%  padding; (4) highlight trigger ranges and boxes must be valid; and (5) highlights within a region must refer to distinct elements.


\paragraph{Generation and validation.}
For each page, the system sends the page image and a structured prompt to Gemini 3 Pro, which jointly generates region decomposition, narration text, and highlight bindings in JSON format. The output is then automatically validated for structure, character ranges, and positive-area bounding boxes. If a region lacks a valid zoom box, the system falls back to a full-page box $(0,0,1,1)$ to preserve playback continuity. Only validated outputs proceed to speech synthesis.
\subsubsection{Speech Synthesis.}
Each region's narration is synthesized with Azure AI Speech Text-to-Speech~\cite{azure_tts_docs}. Audio files and duration metadata are pregenerated and stored for latency-free playback.

\subsection{Playback Engine}

\subsubsection{Sequential Region Playback.}
The playback engine renders the two-level guidance output along the narration timeline. For each region~$r_i$ on the current page, the engine executes three operations in parallel: (1)~zooming the viewport to~$b_i$ (region-level), (2)~playing the pregenerated narration audio for~$n_i$, and (3)~updating the highlight overlay according to the trigger timing in~$\mathcal{H}_i$ (element-level). The system advances to the next region once all three operations complete.

\subsubsection{Viewport Zoom.}
When a region becomes active, the system computes a CSS transform (scale~+ translate) that centers and fits the region's bounding box within the display container. The first region---the full-page overview---displays the entire slide at approximately unit scale, establishing a spatial frame of reference; subsequent regions progressively zoom into local details.

\subsubsection{Narration-Synchronized Highlighting.}
The highlighting mechanism operates at sentence granularity within each region. The narration text is segmented at sentence boundaries, and each segment is allocated a time budget proportional to its share of characters in the narration:
\begin{equation}
  t_j = \frac{|c_j|}{\sum_k |c_k|} \times T_{\text{audio}}
\end{equation}
where $|c_j|$ is the character count of segment~$j$ and $T_{\text{audio}}$ is the region's total audio duration. When a segment's time window becomes active, the system checks whether its character range overlaps with any highlight trigger. Matching highlights are rendered as semi-transparent overlay boxes on top of the slide image, positioned using the same coordinate transform as the viewport zoom. When a segment ends, the overlay boxes fade out with a brief transition.

This mechanism ensures that when the narration says, for example, ``this external microphone on the outside picks up surrounding noise,'' the corresponding visual elements on the slide are highlighted at precisely that moment---reproducing the temporal coordination of a teacher pointing at the board while speaking.



\subsection{System Quality and Robustness Evaluation}

We evaluated generation quality in two stages. We first assessed every guidance item in the materials used in the user study, and then examined highlight-localization robustness on a broader and more diverse lecture-slide collection.

\subsubsection{Quality on Study Materials}

Two independent raters watched the system-generated Dynamic Visual Guidance videos alongside the original slide images and scored every zoom and highlight item on each page of the two study decks (noise-canceling headphones and MRI principles, four pages each). Five aspects were rated on a 0--3 scale (0~= completely incorrect, 3~= precisely accurate): zoom position, zoom trigger timing, highlight position, highlight trigger timing, and region narration order.

In total, the evaluation covered 38 zoom items, 46 highlight items, and 8 region-order items, yielding 176 paired judgments. The raters agreed exactly on 74.4\% of the judgments and within $\pm 1$~point on 99.4\%. Four of the five dimensions were rated at or near ceiling, and both raters assigned the maximum score to every zoom-timing item. Because the ratings were highly concentrated at the ceiling and the zoom-timing dimension had zero variance, a single chance-corrected agreement coefficient would be difficult to interpret. We therefore report exact and within-one-point agreement rather than Cohen's $\kappa$ for this evaluation.

\begin{table}[t]
  \centering
  \small
  \caption{Dynamic Visual Guidance generation quality on the study materials. Means and standard deviations are calculated from the pooled ratings of two independent raters.}
  \label{tab:quality}
  \begin{tabular}{lcc}
    \toprule
    Dimension & Mean (0--3) & SD \\
    \midrule
    Zoom position accuracy      & 2.71 & 0.46 \\
    Zoom trigger timing         & 3.00 & 0.00 \\
    Highlight position accuracy & 2.36 & 0.69 \\
    Highlight trigger timing    & 2.91 & 0.28 \\
    Region narration order      & 2.94 & 0.25 \\
    \bottomrule
  \end{tabular}
\end{table}

As shown in Table~\ref{tab:quality}, zoom trigger timing, highlight trigger timing, and region narration order approached ceiling, indicating that the generated guidance was well synchronized with the narration and followed a natural instructional sequence. Zoom position was also rated highly, indicating that the intended region was generally visible with sufficient surrounding context.

\begin{figure}[t]
  \centering
  \includegraphics[width=\columnwidth]{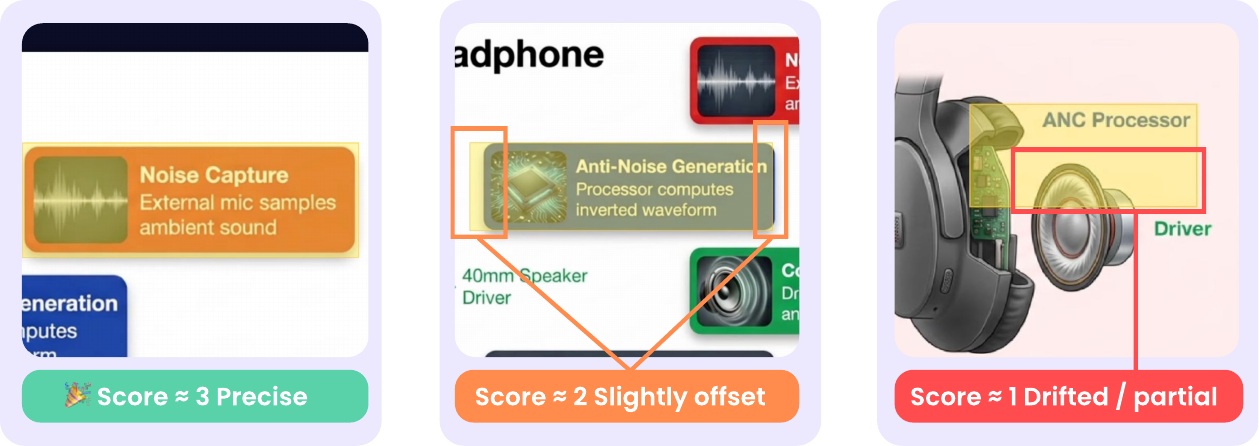}
  \caption{Representative examples of highlight position accuracy. Score~= 3: precisely aligned. Score~= 2: slightly offset but still unambiguous. Score~= 1: partially or ambiguously localized.}
  \Description{Three side-by-side slide crops, each showing a semi-transparent highlight overlay. The first is precisely aligned with the intended target, the second is slightly offset but still unambiguously identifies the target, and the third partially covers the intended target or extends toward an adjacent element.}
  \label{fig:highlight-quality}
\end{figure}

Highlight position was the lowest-scoring dimension ($M = 2.36$, $SD = 0.69$), although 88.0\% of the pooled ratings received a score of 2 or 3. Figure~\ref{fig:highlight-quality} shows representative examples at different score levels. A score of 2 indicates that the intended referent was correctly identified with a minor spatial offset. A score of 1 indicates partial or ambiguous localization, whereas a score of 0 indicates localization to unrelated content.

\subsubsection{Robustness Across Slide Components}

\begin{figure*}[t]
  \centering
  \includegraphics[width=\textwidth]{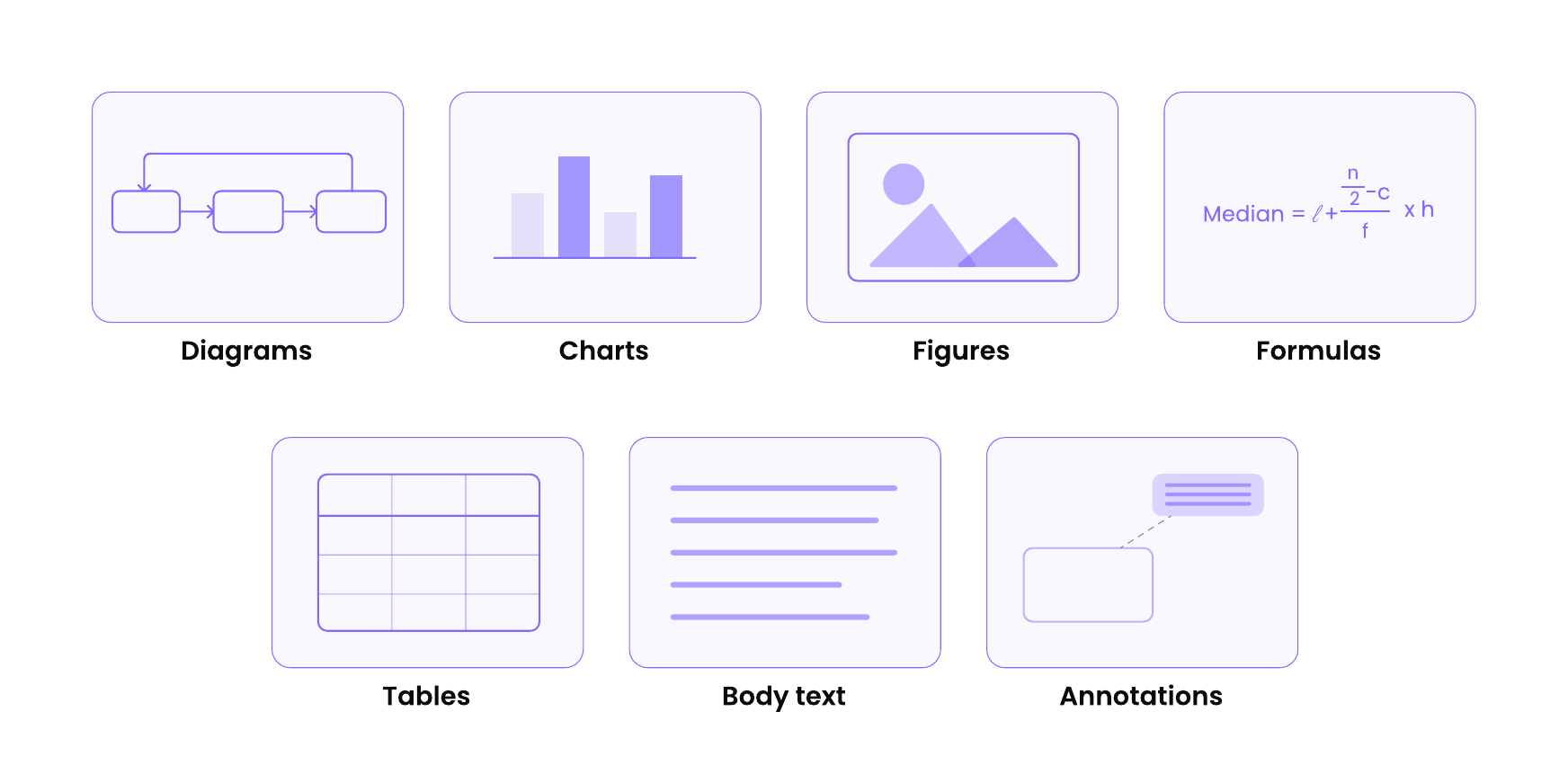}
  \caption{The seven slide component types a highlight can target. Each panel shows a schematic of the component's typical visual form.}
  \Description{Seven labeled panels arranged in two rows, each showing a cropped region of a lecture slide with a semi-transparent highlight overlay: a diagram, a chart, a figure, a formula, a table, a body-text block, and an annotation.}
  \label{fig:components}
\end{figure*}

Lecture slides vary widely in visual style, but the elements a highlight can target are more constrained. Highlights in our collection targeted seven component types: diagrams, charts, figures, formulas, tables, body text, and annotations, illustrated in Figure~\ref{fig:components}.

Because highlight position was the weakest dimension in the study-material evaluation, we examined its robustness on a broader collection. We sampled four slides from each of 20 publicly available lecture decks drawn from 13 universities, yielding 80 slides and 428 highlight items spanning all seven component types. Two independent raters scored every highlight using the same 0--3 positional rubric. This evaluation focuses on the localization accuracy of generated highlights rather than the completeness of highlight coverage.

Across the 856 pooled judgments, highlight position received a mean score of 2.65 ($SD = 0.64$). A total of 91.2\% received a score of 2 or 3, comparable to the 88.0\% observed on the study materials. The remaining 8.8\% received a score of 1, and no judgment received a score of 0. Inter-rater agreement was 82.7\% exact, with a quadratic-weighted Cohen's $\kappa$ of .75.

Together, the two evaluations indicate that the highlight-localization component identifies the intended referent across a range of slide component types. Residual inaccuracies primarily involved incomplete highlighting of the intended referent or minor spatial ambiguity rather than misdirection to unrelated content.
\section{User Study}
To evaluate whether Dynamic Visual Guidance reduces the split-attention effect in narrated slide viewing, we conducted a within-subjects study ($N = 24$) comparing the Static condition against the Dynamic Visual Guidance condition with identical narration content.

\subsection{Study Design}
We used a within-subjects design~\cite{lazar2017research} with two conditions: the \textbf{Static condition}, in which the slide was displayed as a full, still image with AI-generated narration audio, and the \textbf{Dynamic Visual Guidance condition}, in which the same narration was accompanied by automated region zoom and narration-synchronized highlights. The two conditions used identical slide content and the same pre-generated narration audio (5~minutes 28~seconds per video), differing only in whether visual guidance was rendered during playback.

The research team developed two custom instructional slide decks: one on MRI principles (4~pages) and one on noise-canceling headphone technology (4~pages). To verify that the two topics were comparable in difficulty, we conducted a preliminary evaluation in which four participants viewed both slide sets under the Static condition and completed the same comprehension tests. Average scores were closely matched across topics (MRI: $M = 7.25$, $SD = 2.22$; Headphones: $M = 7.00$, $SD = 2.16$; out of 13), suggesting that the materials were sufficiently similar in difficulty for counterbalanced assignment.

To control for order effects and topic-to-condition assignment, we used a counterbalanced presentation scheme~\cite{lazar2017research}. Each participant experienced both conditions and both topics exactly once, with the order of conditions, order of topics, and topic-condition pairing balanced across four presentation orders.

\subsection{Research Questions}
This study is guided by the following research questions:
\begin{itemize}
    \item \textbf{RQ1:} Does \emph{Dynamic Visual Guidance} reduce the split-attention effect and perceived cognitive load compared to static slides with identical narration?
    \item \textbf{RQ2:} How do viewers perceive the overall experience of \emph{Dynamic Visual Guidance} in terms of usability and user experience?
    \item \textbf{RQ3:} Does \emph{Dynamic Visual Guidance} improve immediate comprehension of narrated slide content compared to static slides with identical narration?
\end{itemize}

\subsection{Participants}
We recruited 24 participants (16 female, 8 male; aged 18--29, $M = 22.1$, $SD = 3.1$) from a local university. Sixteen were undergraduates, and eight held or were pursuing a postgraduate degree.
Self-reported English reading proficiency averaged 5.0 ($SD = 1.3$) on a 7-point scale. Prior familiarity was low for both topics (MRI: $M = 2.2$, $SD = 1.5$; headphones: $M = 2.3$, $SD = 1.4$; 7-point scales). Each participant received HK\$100 in supermarket vouchers as compensation. The study protocol was approved by the institutional review board of our institution.

\subsection{Procedure}
The study comprised two sessions: an in-person session (approximately 35~minutes) and an online follow-up 24~hours later (approximately 8~minutes). All in-person sessions were conducted individually. Participants used their own smartphones (screen sizes ranged from 6.1 to 6.7~inches).

\subsubsection{In-Person Session.}
Participants completed a background questionnaire collecting demographic information and self-rated topic familiarity. They then completed two learning rounds in the order assigned by their counterbalancing group.

In each round, participants watched a 5-minute-28-second instructional video on their phone, followed immediately by a comprehension test, a perceived attentional guidance questionnaire, a workload questionnaire (NASA-TLX), and a user experience questionnaire (UEQ-S). A one-minute break separated the two rounds. After both rounds, participants completed a forced-choice preference comparison and a semi-structured interview (approximately 15~minutes, audio-recorded with consent).

\subsubsection{Delayed Session.}
Twenty-four hours after the in-person session, participants received an online questionnaire containing a delayed comprehension test covering both topics, designed to assess retention and transfer.

\subsection{Measures}

\subsubsection{Workload.}
Because split attention increases cognitive load through the additional demands of visual search and cross-referencing~\cite{ayres_split-attention_2014}, we measured perceived workload using the NASA-TLX on a 1--10 scale~\cite{hart_development_1988}.

\subsubsection{Perceived Attentional Guidance.}
To capture proximal experiential indicators of reduced split attention, such as knowing where to look and perceiving narration--visual correspondence, we assessed perceived attentional guidance during viewing using six 7-point Likert items (full item text in Appendix~A).

\subsubsection{User Experience.}
To assess overall user experience, we used the UEQ-S~\cite{schrepp_design_2017}, which measures both pragmatic quality and hedonic quality. This allowed us to evaluate whether Dynamic Visual Guidance was perceived as a more usable, supportive, and engaging viewing format than static slides.

\subsubsection{Comprehension.}
Because split attention impairs how well viewers can follow and retain spatially separated information~\cite{chandler_split-attention_1992}, we assessed comprehension through both immediate and delayed tests, following the common distinction between retention and transfer outcomes~\cite{mayer2009multimedia}. Each topic included an immediate test (3 multiple-choice items at 1 point each and 2 short-answer items at 3 points each; maximum 9 points) and a 24-hour delayed test. The delayed test comprised two retention questions and two transfer questions per topic. Each subscore (retention and transfer) contained one multiple-choice question (1 point) and one short-answer question (3 points), yielding a maximum of 4 points per subscore and 8 points per topic. Short-answer items were scored independently by two raters, with disagreements resolved by a third rater.

\subsubsection{Preference.}
To capture participants' overall comparative judgment of the two formats, four forced-choice questions asked which condition was easier to follow, better supported understanding, felt less demanding, and was preferred for future use.

\subsection{Qualitative Data Collection}

\subsubsection{Semi-Structured Interview.}
After completing both learning conditions, participants took part in a semi-structured interview about their overall comparisons between conditions, their experiences with zoom and highlight guidance, and mobile-specific considerations. Interviews were audio-recorded with consent, transcribed, and later analyzed using thematic analysis~\cite{braun_using_2006}.

\subsection{Analysis}

For all paired comparisons between conditions, we first tested the normality of the difference scores using the Shapiro--Wilk test~\cite{shapiro1965analysis} ($\alpha = .05$). Measures that satisfied normality were analyzed with paired $t$-tests, with effect sizes reported as Cohen's~$d$~\cite{cohen1988statistical}; measures that violated normality were analyzed with Wilcoxon signed-rank tests~\cite{wilcoxon1945individual}, with effect sizes reported as $r = Z / \sqrt{N}$. For parametric comparisons, we report 95\% confidence intervals for the mean paired difference. Within each outcome family, we designated a single primary measure tested at $\alpha = .05$ without further adjustment; item-level breakdowns were treated as exploratory analyses with Holm-corrected $p$-values~\cite{holm1979simple}. Internal consistency of multi-item scales was assessed with Cronbach's~$\alpha$~\cite{cronbach1951coefficient}. Preference responses were analyzed using exact binomial tests, excluding tied responses (``no difference''). Interview data were analyzed using thematic analysis~\cite{braun_using_2006}: transcripts were segmented into semantic units, then two researchers independently coded all units. A third researcher arbitrated disagreements to produce the final codebook of 68 codes.
\begin{figure*}[t]
  \centering
  \includegraphics[width=\textwidth]{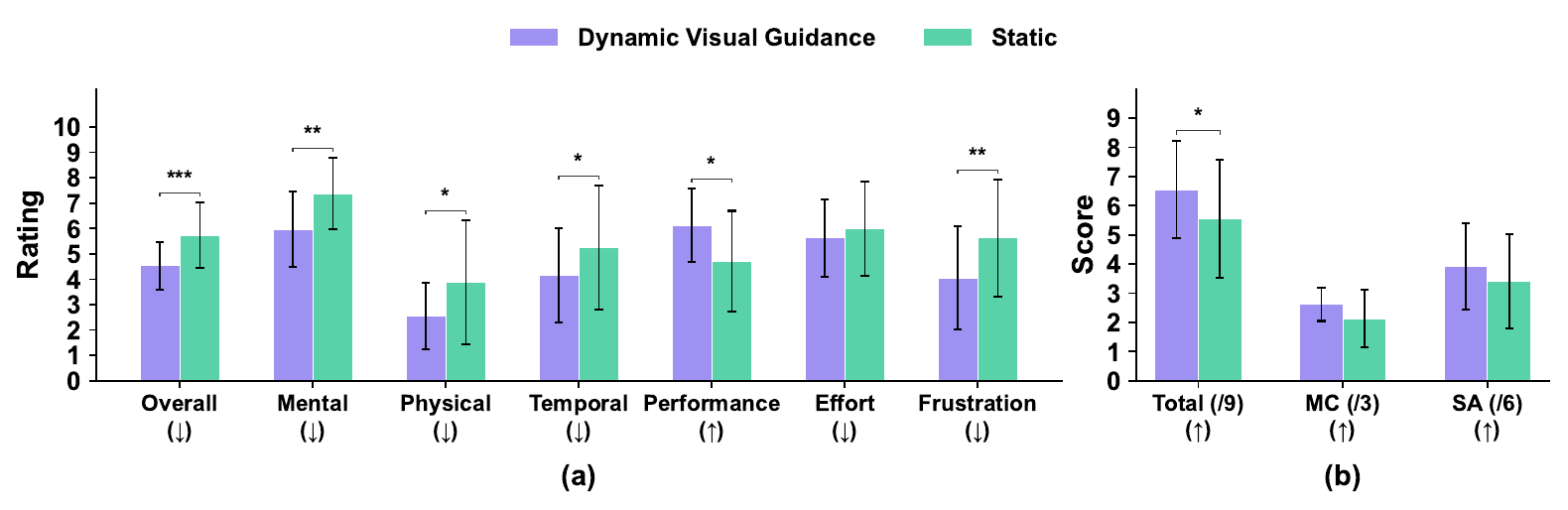}
  \caption{Workload and immediate comprehension by condition ($N = 24$), addressing RQ1 and RQ3. (a) NASA-TLX ratings (1--10): lower indicates less workload except Performance~($\uparrow$). The overall score is the primary measure (with Performance reverse-scored); dimensions are exploratory. (b) Immediate comprehension scores: Total (9 points) is the primary measure; MC (3 points) and SA (6 points) are exploratory. Bars show means $\pm 1$ SD. Physical used a Wilcoxon signed-rank test; all other comparisons used paired $t$-tests (* $p < .05$, ** $p < .01$, *** $p < .001$).}
  
  \Description{Two grouped bar charts side by side (left panel wider), comparing Dynamic Visual Guidance (purple) and Static (green). Panel (a) shows the NASA-TLX overall score and six dimensions---Mental, Physical, Temporal, Performance, Effort, and Frustration---with most workload dimensions favoring dynamic and higher Performance ratings for dynamic. Panel (b) shows immediate comprehension scores for Total, MC, and SA, with a significance marker on Total.}
  \label{fig:workload_comprehension}
\end{figure*}

\begin{figure*}[t]
  \centering
  \includegraphics[width=\textwidth]{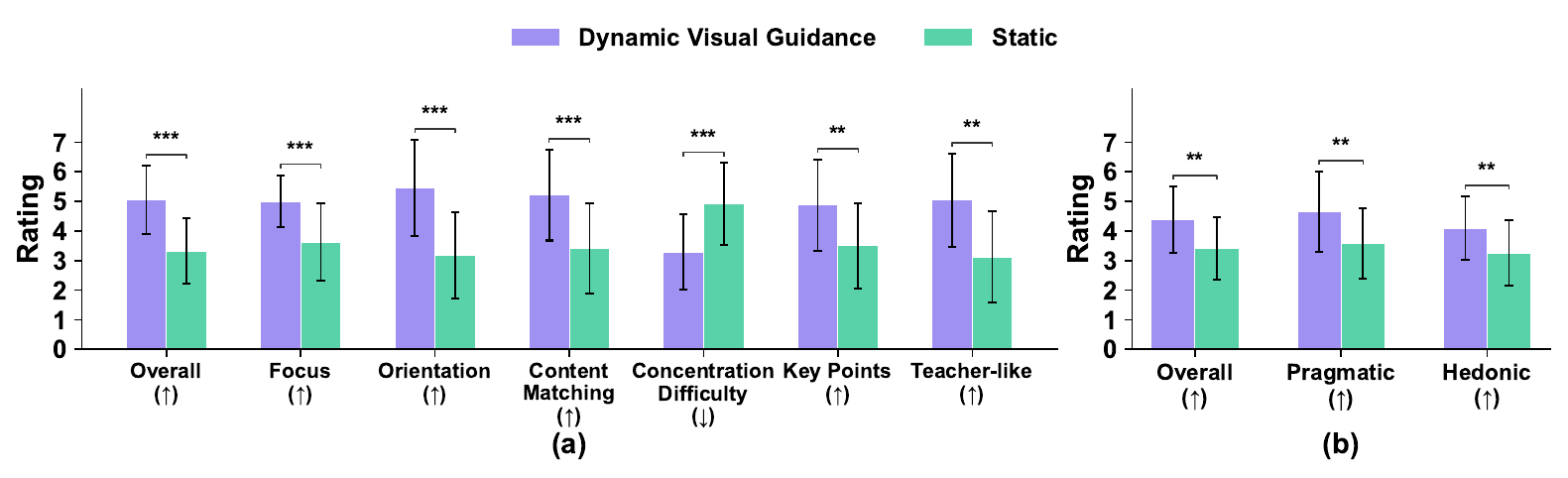}

  \caption{Perceived attentional guidance and user experience by condition ($N = 24$), addressing RQ1 and RQ2. (a) Attentional-guidance ratings (1--7): the overall score is the primary measure; item-level results are exploratory, with Concentration Difficulty reverse-coded. (b) UEQ-S ratings (1--7): the overall score is the primary measure; Pragmatic and Hedonic are exploratory. Bars show means $\pm 1$ SD. All comparisons used paired $t$-tests (* $p < .05$, ** $p < .01$, *** $p < .001$).}

  \Description{Two grouped bar charts side by side (left panel wider), comparing Dynamic Visual Guidance (purple) and Static (green). Panel (a) shows the overall score plus six perceived attentional-guidance items---Focus, Orientation, Content Matching, Concentration Difficulty (reverse-coded), Key Points, and Teacher-like---with all ratings favoring dynamic. Panel (b) shows UEQ-S Overall, Pragmatic, and Hedonic, all favoring dynamic and marked as statistically significant.}
  \label{fig:guidance_ueq}
\end{figure*}

\section{Results}

We report results by outcome family below, following the analysis plan described in Section~5.7. Unless otherwise noted, values are reported as $M$~($SD$). For each family, we present the pre-specified primary measure first, followed by exploratory item-level breakdowns where relevant.

\subsection{Workload and Perceived Attentional Guidance}

\subsubsection{Workload.}
The Dynamic Visual Guidance condition significantly reduced perceived workload relative to the Static condition. We computed an overall workload score from the NASA-TLX as the unweighted mean of all six items (with Performance reverse-scored so that lower values consistently indicate lower workload). The overall score was significantly lower in the Dynamic Visual Guidance condition ($M = 4.53$, $SD = 0.94$) than in the Static condition ($M = 5.74$, $SD = 1.29$; $\Delta = -1.20$, 95\% CI $[-1.80, -0.60]$, $t(23) = -4.14$, $p < .001$, $d = 0.84$), indicating a large reduction in perceived workload.

At the dimension level, five of six dimensions remained significant after Holm correction (Figure~\ref{fig:workload_comprehension}, left panel). Mental Demand showed the strongest effect ($\Delta = -1.42$, 95\% CI $[-2.11, -0.72]$, $p_\text{H} = .002$, $d = 0.86$), suggesting that the guidance primarily relieved the effort of locating relevant content rather than requiring viewers to determine it themselves. Frustration was also substantially lower ($\Delta = -1.58$, $p_\text{H} = .008$, $d = 0.73$), consistent with the interpretation that reducing spatial uncertainty made the experience easier to follow. Performance self-assessment was significantly higher in the Dynamic Visual Guidance condition ($p_\text{H} = .023$, $d = 0.62$). Effort was the only dimension that did not reach significance ($p = .205$), indicating that the clearest workload benefits were concentrated in search- and uncertainty-related dimensions.

\subsubsection{Perceived Attentional Guidance.}
Perceived attentional-guidance ratings strongly favored the Dynamic Visual Guidance condition. The six items showed good internal consistency (Cronbach's $\alpha = .90$ for the Dynamic Visual Guidance condition and $.86$ for the Static condition, with Q4 reverse-coded). The overall score was substantially higher in the Dynamic Visual Guidance condition ($M = 5.05$, $SD = 1.16$) than in the Static condition ($M = 3.32$, $SD = 1.11$; $\Delta = 1.73$, 95\% CI $[1.04, 2.42]$, $t(23) = 5.19$, $p < .001$, $d = 1.06$), indicating a large overall improvement in perceived attentional guidance.

All six individual items remained significant after Holm correction (Figure~\ref{fig:guidance_ueq}, left panel). The largest effect appeared on Q2 (``I knew where to look at each moment''; $\Delta = 2.29$, 95\% CI $[1.33, 3.25]$, $p_\text{H} < .001$, $d = 1.01$), directly reflecting the system's core design objective of reducing spatial uncertainty during narrated slide viewing. Participants also reported feeling more focused (Q1; $d = 0.90$), stronger narration--visual correspondence (Q3; $d = 0.91$), easier identification of key points (Q5; $d = 0.70$), and a more teacher-like attentional-guidance experience (Q6; $d = 0.79$). The reverse-coded Q4 (difficulty concentrating) was significantly lower in the Dynamic Visual Guidance condition ($\Delta = -1.62$, $p_\text{H} < .001$, $d = 0.99$). Together, these item-level effects indicate that the Dynamic Visual Guidance condition improved spatial orientation, narration--visual coordination, and perceived instructor-like attentional support.

\subsection{User Experience}

The Dynamic Visual Guidance condition also improved overall user experience. Internal consistency was high for both subscales across conditions (Pragmatic: $\alpha = .88$; Hedonic: $\alpha = .84$--.85). Overall user experience, measured with the UEQ-S, was significantly higher in the Dynamic Visual Guidance condition ($M = 4.38$, $SD = 1.13$) than in the Static condition ($M = 3.41$, $SD = 1.05$; $\Delta = 0.96$, 95\% CI $[0.36, 1.56]$, $t(23) = 3.33$, $p = .003$, $d = 0.68$; Figure~\ref{fig:guidance_ueq}, right panel), indicating a medium-to-large improvement in the overall perceived quality of the viewing experience. Exploratory breakdowns confirmed this advantage for both pragmatic quality ($\Delta = 1.08$, $p_\text{H} = .008$, $d = 0.69$) and hedonic quality ($\Delta = 0.84$, $p_\text{H} = .009$, $d = 0.58$), both remaining significant after Holm correction. This suggests that the Dynamic Visual Guidance condition was experienced as both more supportive and more engaging.

\subsection{Comprehension}

Short-answer questions on both immediate and delayed tests were independently scored by two raters following a predefined rubric (0--3 points per question). Across all 192 ratings, inter-rater reliability was high: exact agreement reached 94.8\%, agreement within $\pm 1$~point was 100\%, and Cohen's $\kappa$ (quadratic-weighted)~\cite{cohen1968weighted} was .970 (immediate) and .980 (delayed). All ten disagreements differed by one point and were resolved by a third independent rater.

The Dynamic Visual Guidance condition also produced a significant advantage on the primary comprehension outcome. Participants scored significantly higher overall in the Dynamic Visual Guidance condition ($M = 6.54$, $SD = 1.67$) than in the Static condition ($M = 5.54$, $SD = 2.02$), representing about an 18\% increase relative to the Static condition after a single 5.5-minute viewing session ($\Delta = 1.00$, 95\% CI $[0.08, 1.92]$, $t(23) = 2.24$, $p = .035$, $d = 0.46$; Figure~\ref{fig:workload_comprehension}, right panel). This pattern suggests that the Dynamic Visual Guidance condition improved viewers' understanding of narrated content rather than merely making the experience feel easier.

Exploratory breakdowns showed that the gain was distributed across both multiple-choice items ($\Delta = 0.50$, $p_\text{H} = .105$, $d = 0.42$) and short-answer items ($\Delta = 0.50$, $p_\text{H} = .218$, $d = 0.26$), rather than being driven by only one response format; however, neither subscore reached significance after Holm correction. Taken together, the immediate results indicate a modest but reliable advantage for first-pass understanding in the Dynamic Visual Guidance condition.

\paragraph{Delayed Comprehension (24-Hour Follow-Up).}
Twenty-four hours after the in-person session, participants completed a delayed test comprising two retention questions and two transfer questions per topic, for a maximum of 8~points per condition. The overall delayed score showed a directionally consistent advantage for the Dynamic Visual Guidance condition ($M = 5.12$, $SD = 1.60$) over the Static condition ($M = 4.46$, $SD = 2.06$), though this difference did not reach the pre-specified significance threshold ($\Delta = 0.67$, Wilcoxon $W = 49$, $p = .054$, $r = .39$). Thus, the immediate comprehension advantage showed partial persistence after 24~hours, but with weaker statistical support than in the immediate test.

Exploratory breakdowns suggested that this trend was more pronounced for retention items ($\Delta = 0.46$, $p_\text{H} = .155$, $d = 0.38$) than for transfer items ($\Delta = 0.21$, $p_\text{H} = .347$, $d = 0.20$), although neither reached significance after Holm correction. Overall, the delayed pattern was directionally consistent with the immediate result, but weaker and more evident for retention than for transfer.

\subsection{Preference}

\begin{figure}[t]
  \centering
  \includegraphics[width=\columnwidth]{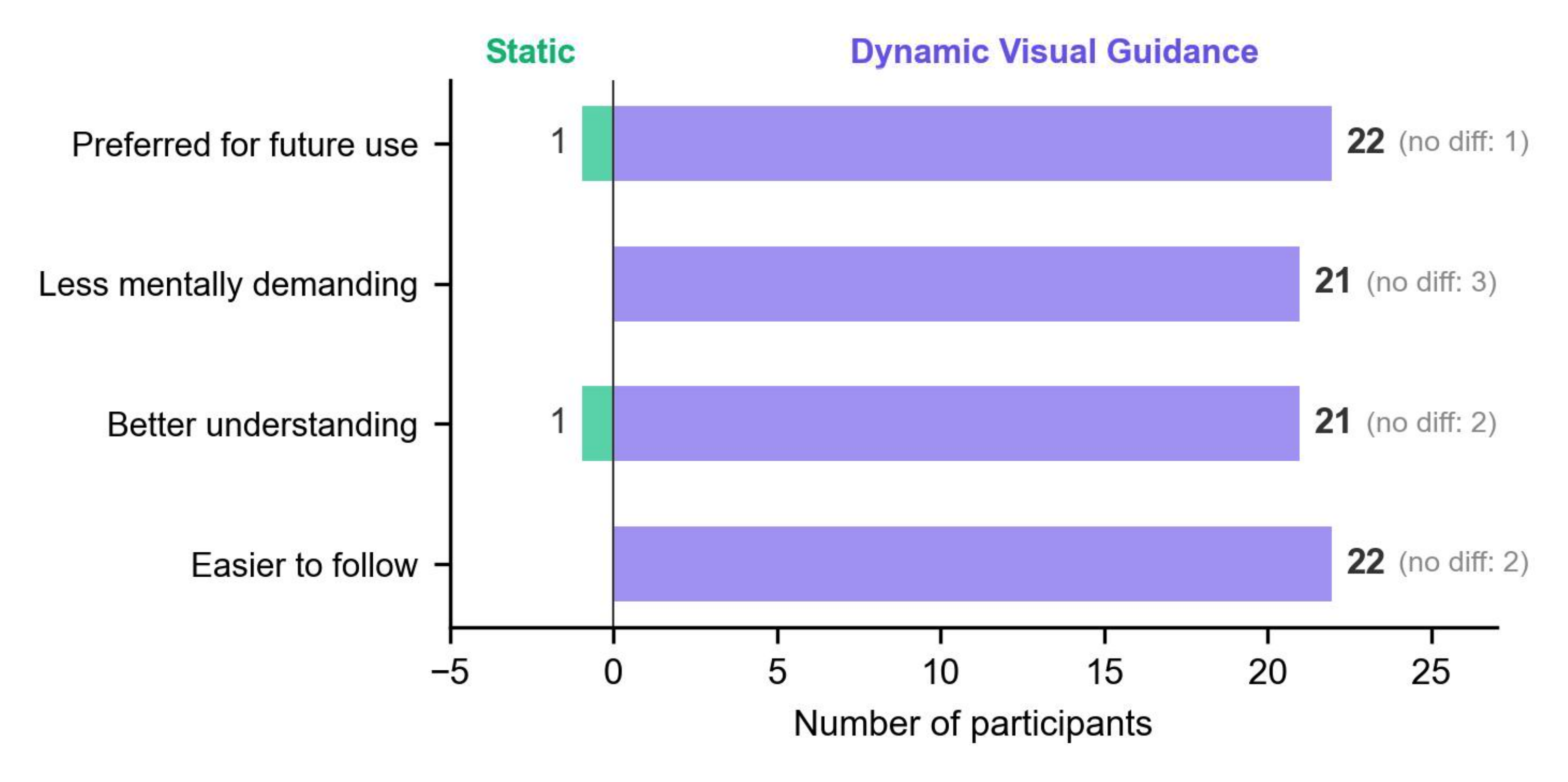}
  \caption{Preference distribution across four forced-choice questions ($N = 24$). Numbers indicate participant counts; remaining reported no difference.}
  \Description{Horizontal diverging bar chart showing 21--22 participants preferring dynamic on all four questions.}
  \label{fig:preference}
\end{figure}

Participants overwhelmingly preferred the Dynamic Visual Guidance condition across all four forced-choice questions (Figure~\ref{fig:preference}). Exact binomial tests were conducted on non-tied responses (excluding participants who reported ``no difference''). All four comparisons were highly significant ($p < .001$), indicating a clear participant preference for the Dynamic Visual Guidance condition.

\subsection{Robustness Checks}

To verify that the counterbalancing design controlled for confounds, we checked for presentation-order and topic effects on comprehension. Participants who experienced the Dynamic Visual Guidance condition in Session~1 showed an identical mean condition advantage ($\Delta = 1.00$, $SD = 2.63$) to those who experienced it in Session~2 ($\Delta = 1.00$, $SD = 1.76$; $t(22) < 0.01$, $p > .99$), suggesting no meaningful order bias. The two instructional topics produced comparable scores (MRI: $M = 6.25$; Headphones: $M = 5.83$; $p = .454$), and the condition advantage did not differ between topics ($\Delta_\text{MRI} = 1.42$, $\Delta_\text{Headphones} = 0.58$; $p = .362$), providing no evidence of a Condition~$\times$~Topic interaction.

\subsection{Interview Findings}

Semi-structured interviews were conducted with all 24 participants immediately after the experiment to help explain how participants experienced the two conditions beyond the quantitative measures. All interviews were conducted in local language; quotes below are translated by the authors. Interviews were transcribed, segmented into participant semantic units ($N = 602$), and analyzed through thematic analysis~\cite{braun_using_2006}. Two researchers independently coded all semantic units; a third researcher arbitrated disagreements, producing 68 agreed codes from which we identified five themes organized around the three research questions.

\subsubsection{Theme 1: Spatial Orientation}
The most prevalent finding was that the Dynamic Visual Guidance condition eliminated the need for visual search. Participants described how the zoom-and-highlight cues directed their gaze to the exact region under discussion (coded 62 times across 23 of 24 participants). One participant explained: \emph{``[With dynamic mode] it shows you where to look---I just follow along''} (F003). Conversely, in the Static condition, participants reported repeatedly losing track of the narrated content on information-dense slides (41 mentions, 20 participants): \emph{``With static, I had no idea which part they were talking about, so I was scanning all over the place''} (F002). Some participants noted that in the Static condition they shifted part of their attention to the subtitle transcript to compensate, whereas in the Dynamic Visual Guidance condition they could attend to the slide content itself.

\subsubsection{Theme 2: Teacher-Like Attentional Guidance}
Forty-three references across 23 participants described the Dynamic Visual Guidance condition as evoking the feeling of a teacher pointing at a whiteboard, suggesting that the system reconstructed the kind of attentional guidance normally provided by a live teacher. Critically, many considered it \emph{superior} to a real teacher's pointer because the zoom provided magnified, focused views: \emph{``A teacher pointing tells you where to look, but the zoom brings the content right in front of you---that's even better''} (F003). Participants also valued the segmented, step-by-step presentation: rather than confronting an entire slide at once, the Dynamic Visual Guidance condition decomposed it into sequential regions that synchronized audio and visual channels (14 mentions of segmented presentation, 8 mentions of audio--visual synchrony).

\subsubsection{Theme 3: Cognitive Benefits}
Participants reported that the Dynamic Visual Guidance condition improved their comprehension (42 mentions, 22 participants) and sustained their focus (20 mentions, 13 participants). Representative comments included: \emph{``I can connect what I hear to what's on the slide, which really helps me understand''} (F005) and \emph{``Because it zooms in, there's less on screen, so I can focus on just those few words''} (F002). In contrast, the Static condition was described as disengaging: \emph{``Static mode felt worse than a regular online lecture---at least a teacher has a mouse cursor''} (F012). However, 15 participants across 43 instances noted that content difficulty moderated the perceived benefit, suggesting that the guidance advantage is most salient for unfamiliar, information-dense material.

\subsubsection{Theme 4: Mobile Amplification}
Thirteen participants reported that the Dynamic Visual Guidance condition's advantages were amplified on small screens, while the Static condition suffered disproportionately: \emph{``On a phone the static text is way too small---I didn't even want to look at it''} (F006). The automatic zoom effectively compensated for limited screen real estate by enlarging the relevant region without manual pinch-to-zoom. Eight participants, however, perceived no meaningful difference between conditions on mobile.

\subsubsection{Theme 5: Future Use and Design Recommendations}
Participants envisioned using the system primarily for previewing unfamiliar material (15 mentions) and reviewing lecture content (19 mentions). The most cited concern was temporary loss of global context when zoomed in (32 mentions, 15 participants), and five participants wished for viewer control over pacing, such as pause and rewind. Design suggestions included adding a persistent minimap to preserve overview context (5 mentions), supporting variable playback speed and pause controls (5 mentions), and varying highlight colors to reduce visual monotony (9 mentions). Notably, four participants independently suggested interactive Q\&A---a feature already implemented in \textsc{GuideSlide} but disabled in the experiment to isolate the effect of visual guidance.

\section{Discussion}

\subsection{Narration-Synchronized Guidance Reduces Split Attention and Cognitive Load}

(RQ1) \emph{Does Dynamic Visual Guidance reduce the split-attention effect and perceived cognitive load compared to static slides with identical narration?} Our results provide clear evidence that it does, at both the macro and micro level of the split-attention burden we identified.

The quantitative results are direct. Overall workload was significantly lower in the Dynamic Visual Guidance condition ($d = 0.84$), with Mental Demand showing the strongest reduction ($d = 0.86$). This is precisely what split-attention theory predicts: when viewers no longer need to search for where the narration refers to, they can dedicate more cognitive resources to actually understanding the content. Our custom perceived attentional guidance scale showed the largest overall effect across all measures ($d = 1.06$), and the single strongest item was ``I knew where to look at each moment'' ($d = 1.01$)---directly reflecting the macro-level spatial orientation that our system was designed to provide.

These effects expose a gap in prior automatic narrated slide systems. AutoLectures~\cite{holmberg2025generatingnarratedlecturevideos} and Mizukami et al.~\cite{mizukami2025study} address only the micro-level component of the signaling principle---marking which specific element to attend to. Neither system provides macro-level spatial framing: they do not direct viewers to the relevant region before cueing specific elements, leaving the first and more fundamental layer of split attention unresolved. Our results suggest this missing layer matters: the largest effects in our study appeared precisely on the measures that reflect spatial orientation and region-level guidance, not element identification alone. To our knowledge, \textsc{GuideSlide} is the first system to address macro-level spatial framing in narrated slide viewing, and the first to provide empirical evidence that doing so meaningfully reduces split attention.

The interview data further support this interpretation. Twenty-three of 24 participants described the Dynamic Visual Guidance condition as eliminating the need for visual search (62 mentions, Theme~1), and many described it as superior to a teacher's pointer because the zoom magnified the referenced content rather than merely indicating its location: \emph{``A teacher pointing tells you where to look, but the zoom brings the content right in front of you''} (F003). Conversely, viewers in the Static condition repeatedly lost track of narrated content (41 mentions, 20 participants), with some shifting attention to the subtitle transcript as a compensatory strategy: \emph{``With static, I had no idea which part they were talking about, so I was scanning all over the place''} (F002).

\subsection{Dynamic Visual Guidance Delivers a Better Viewing Experience}

(RQ2) \emph{How do viewers perceive the overall experience of Dynamic Visual Guidance in terms of usability and user experience?} Our results provide clear evidence that Dynamic Visual Guidance delivers a qualitatively better viewing experience.

The most immediate evidence comes from four forced-choice preference questions: 21--22 of 24 participants preferred the Dynamic Visual Guidance condition across all four (all $p < .001$). UEQ-S~\cite{schrepp_design_2017}, a validated standard scale, confirmed this across both pragmatic quality ($d = 0.69$) and hedonic quality ($d = 0.58$), indicating that viewers found the system both more usable and more engaging. Frustration was also substantially lower in the Dynamic Visual Guidance condition ($d = 0.73$), suggesting that the guidance reduced not only cognitive load but also the affective cost of navigating information-dense slides without support. This advantage did not arise only from making the presentation more visually dynamic, but from reducing the split-attention burden of coordinating narration with spatially distributed slide content.

Interview data further contextualized these effects. Theme~3 captured 20 mentions of sustained focus across 13 participants, and the Static condition was consistently described as actively disengaging---consistent with recent findings that passive viewing without interactive support reduces engagement~\cite{bodonhelyi_passive_2025}. One participant drew directly on their own experience preparing slides: \emph{``When I did a pitch deck without animations, the professor said he couldn't tell which part I was talking about---that's exactly why I think this dynamic mode works''} (F023). This mirrors the split-attention problem our system was designed to solve.

Prior work on automatic narrated slide generation has relied exclusively on technical evaluation metrics, explicitly acknowledging the absence of empirical evaluation with real viewers~\cite{holmberg2025generatingnarratedlecturevideos, mizukami2025study}. Our results fill this gap, providing the first user evidence that narration-synchronized hierarchical guidance produces a meaningfully better viewing experience than static slides.

\subsection{Dynamic Visual Guidance Improves Comprehension}

(RQ3) \emph{Does Dynamic Visual Guidance improve immediate comprehension of narrated slide content compared to static slides with identical narration?} Our results show that Dynamic Visual Guidance improved comprehension, with effects evident immediately after viewing and partially persisting at 24-hour follow-up. Viewers scored significantly higher immediately after viewing ($d = 0.46$), and this advantage remained directionally consistent at the 24-hour follow-up, particularly for retention items ($d = 0.38$), though transfer items showed a smaller, non-significant effect ($d = 0.20$). These results are consistent with the signaling principle's prediction that visual cues reduce the integration cost of spatially separated information~\cite{mayer_past_2024, de_koning_towards_2009}: by alleviating split attention at the point of first contact with material, the guidance strengthened initial understanding of narrated content.

The pattern suggests that the primary mechanism is improved initial understanding rather than deeper conceptual restructuring. In this sense, Dynamic Visual Guidance appears to improve comprehension primarily by reducing the attentional coordination demands imposed by split attention, rather than by changing the conceptual depth of the material itself. This is consistent with the nature of the intervention: Dynamic Visual Guidance resolves an attention allocation problem, directing viewers to the right place at the right time, but does not itself prompt active recall or application. Interview data further contextualized this finding: fifteen participants noted that the guidance benefit was most salient for unfamiliar, information-dense material, and participants identified three primary use cases: previewing unfamiliar content (15 mentions), reviewing lecture material (19 mentions), and sharing slide decks without a live presenter. These findings are consistent with the scope of our intervention: \textsc{GuideSlide} is designed to resolve an attention allocation problem, not to deepen conceptual learning itself.

\subsection{Limitations and Future Work}

Our evidence for reduced split attention relies primarily on self-report and performance measures. Future work should incorporate eye-tracking to directly confirm whether the guidance reduces fixation dispersion and increases time-on-target. A confound worth noting is that some participants in the Static condition reported shifting attention to the subtitle transcript; if static scores partially reflect transcript reading, the observed comprehension gain may underestimate the true effect.

Our sample of 24 university students from a single institution limits generalizability. Although the full \textsc{GuideSlide} system supports runtime playback controls, these were intentionally not exposed in the experiment to isolate the effect of Dynamic Visual Guidance itself. Several directions remain open: whether repeated viewing produces cumulative benefits, whether combining guidance with the interactive Q\&A feature already implemented in \textsc{GuideSlide} extends benefits further, and whether the system generalizes beyond information-dense slides with clearly delineated visual regions. Improving VLM spatial grounding could further reduce the remaining localization inaccuracies observed across both evaluation sets: 12.0\% of pooled highlight-position judgments on the study materials and 8.8\% on the broader lecture-slide collection received scores below 2. More broadly, \textsc{GuideSlide} points toward a class of systems that automatically reconstruct teacher-like attentional guidance for any slide-based content---a direction with implications for self-paced learning, professional communication, and any context where slides must be understood without a live presenter.
\section{Conclusion}
We presented \textsc{GuideSlide}, a system that automatically generates narration-synchronized \emph{Dynamic Visual Guidance} for slide presentations without manual annotation. The core mechanism is a hierarchical attention guidance scheme: region-level zoom narrows the visual search space (macro guidance), while element-level highlights pinpoint the specific referent within the zoomed region (micro guidance), both synchronized with the narration timeline. To our knowledge, \textsc{GuideSlide} is the first system to address macro-level spatial framing in narrated slide viewing---an aspect absent from prior automatic narrated slide systems~\cite{holmberg2025generatingnarratedlecturevideos, mizukami2025study}.

A within-subjects study with 24 participants demonstrated that Dynamic Visual Guidance substantially reduced perceived cognitive load, with the largest reduction in mental demand, consistent with the claim that narration-synchronized spatial guidance alleviates split attention during slide viewing (RQ1). Participants rated the Dynamic Visual Guidance condition significantly higher across usability and hedonic quality, with 21--22 of 24 preferring it across all forced-choice comparisons (RQ2). Immediate comprehension improved significantly, with benefits partially persisting at 24-hour follow-up for retention items (RQ3). Qualitative analysis revealed that participants experienced the guidance as teacher-like attentional direction, with many considering it superior to a real teacher's pointer due to its magnification capability. These results establish that resolving the macro-level spatial component of split attention is essential for effective narrated slide viewing.

These findings establish narration-synchronized hierarchical visual guidance as an effective design strategy for slide-based presentations. More broadly, slides are one of the most widely used formats for structured communication, yet their effectiveness has always depended on the presence of a live presenter to direct attention. \textsc{GuideSlide} demonstrates that this dependency can be lifted: by automatically reconstructing teacher-like attentional guidance, it enables slides to communicate as effectively without a presenter as with one---a shift with implications for professional communication, self-paced learning, and any context where visual content must be understood at a distance.

\bibliographystyle{ACM-Reference-Format}
\bibliography{reference}

\clearpage
\appendix
\section*{Appendix}
\addcontentsline{toc}{section}{Appendix}

\section{Perceived Attentional Guidance Questionnaire}
\label{sec:appendix_guidance_questionnaire}

Participants rated each item on a 7-point Likert scale from 1 (\textit{Strongly Disagree}) to 7 (\textit{Strongly Agree}). The figure labels used in Figure~\ref{fig:guidance_ueq} are listed alongside each item below. Item Q4 was reverse-coded in the analysis and is shown as Concentration Difficulty (R) in the figure.

\begin{table}[h]
\centering
\footnotesize
\renewcommand{\arraystretch}{1.15}
\caption{Perceived Attentional Guidance Questionnaire used in the study.}
\label{tab:guidance_questionnaire}
\begin{tabularx}{0.95\textwidth}{@{}c l X@{}}
\toprule
\textbf{Item} & \textbf{Figure label} & \textbf{Statement} \\
\midrule
Q1 & Focus & This presentation style helped me stay focused throughout the explanation. \\
Q2 & Orientation & During the explanation, I knew where to look on the slide at each moment. \\
Q3 & Content Matching & This presentation style made it easy to connect the narration with the corresponding areas on the slide. \\
Q4 & Concentration Difficulty (R) & During the explanation, the presentation style sometimes made it difficult to concentrate on the narrated content. \\
Q5 & Key Points & This explanation method helped me identify the key points. \\
Q6 & Teacher-like & This felt like having a teacher pointing at the slide while explaining. \\
\bottomrule
\end{tabularx}
\end{table}

\section{AI Usage Declaration}
\label{sec:appendix_ai_usage}

This work used AI tools for supportive, non-creative tasks only:
\begin{enumerate}
    \item Literature summarization and related-work organization: GPT-5.4.
    \item Code generation and debugging: Opus 4.6.
\end{enumerate}
All core contributions, research design, analysis, interpretation, and conclusions are solely authored by the researchers.

\section{Counterbalancing Scheme}
\label{sec:appendix_counterbalancing}

The study used a counterbalanced within-subjects design in which each participant experienced both presentation conditions exactly once and both instructional topics exactly once. To control for order effects and topic-to-condition assignment, participants were distributed evenly across four presentation orders.

\begin{table}[h]
\centering
\small
\renewcommand{\arraystretch}{1.15}
\caption{Counterbalancing scheme used in the study.}
\label{tab:appendix_counterbalancing}
\begin{tabularx}{0.95\textwidth}{@{}c X X@{}}
\toprule
\textbf{Group} & \textbf{Session 1} & \textbf{Session 2} \\
\midrule
1 & MRI + Dynamic Visual Guidance & Headphones + Static \\
2 & MRI + Static & Headphones + Dynamic Visual Guidance \\
3 & Headphones + Dynamic Visual Guidance & MRI + Static \\
4 & Headphones + Static & MRI + Dynamic Visual Guidance \\
\bottomrule
\end{tabularx}
\end{table}

\end{document}